\documentclass[11pt]{article}
\usepackage{graphicx}
\usepackage[margin=1.25in]{geometry}
\usepackage[usenames,dvipsnames]{color}
\usepackage{url}
\usepackage{cite}

\usepackage[colorlinks = true,
            linkcolor = blue,
            urlcolor  = blue,
            citecolor = blue,
            anchorcolor = blue]{hyperref}

\def\Title#1{\begin{center} {\LARGE #1 } \end{center}}
\def\Author#1{\begin{center}{ \sc #1} \end{center}}
\def\Address#1{\begin{center}{ \it #1} \end{center}}

\newenvironment{Abstract}{\begin{quotation} \begin{center}
                       ABSTRACT
     \end{center}\bigskip  }{\end{quotation}}

\def\beq{\begin{equation}}
\def\eeq#1{\label{#1}\end{equation}}
\def\eeqn{\end{equation}}

\newenvironment{Eqnarray}%
   {\arraycolsep 0.14em\begin{eqnarray}}{\end{eqnarray}}
\def\beqa{\begin{Eqnarray}}
\def\eeqa#1{\label{#1}\end{Eqnarray}}
\def\eeqan{\end{Eqnarray}}

\let\bar=\overbar

\def\lsim{\mathrel{\raise.3ex\hbox{$<$\kern-.75em\lower1ex\hbox{$\sim$}}}}
\def\gsim{\mathrel{\raise.3ex\hbox{$>$\kern-.75em\lower1ex\hbox{$\sim$}}}}

\def\del{\partial}
\def\Dslash{\not{\hbox{\kern-4pt $D$}}}
\def\dslash{\not{\hbox{\kern-2pt $\del$}}}
\def\pslash{\not{\hbox{\kern-2pt $p$}}}
\def\ETmiss{\not{\hbox{\kern-4pt $E$}}_T}

\def\Dlr{\mathrel{\raise1.5ex\hbox{$\leftrightarrow$\kern-1em\lower1.5ex\hbox{$D$}}}}

\def\MSB{{\bar{M \kern -2pt S}}}
\def\msb{{\bar{\scriptsize M \kern -1pt S}}}

\def\drb{{\bar{\scriptsize D \kern -1pt R}}}

\newcommand\snowmass{\begin{center}\rule[-0.2in]{\hsize}{0.01in}\\\rule{\hsize}{0.01in}\\
\vskip 0.1in Submitted to the  Proceedings of the US Community Study\\ 
on the Future of Particle Physics (Snowmass 2021)\\ 
\rule{\hsize}{0.01in}\\\rule[+0.2in]{\hsize}{0.01in} \end{center}}

\begin{document}
\snowmass


\Title{The Ion Fluorescence Chamber (IFC): \\ A new concept for directional dark matter and topologically imaging neutrinoless double beta decay searches}

\bigskip 

\Author{B.~J.~P.~Jones$^1$, F.~W.~Foss$^2$, J.~A.~Asaadi$^1$, E.~D.~Church$^3$, J.~deLeon$^1$, E.~Gramellini$^4$, O.~H.~Seidel$^1$, T.~T.~Vuong$^2$}

\Address{$^1$Department of Physics, University of Texas at Arlington,
Arlington, Texas 76019, USA\\
$^2$Department of Chemistry and Biochemistry, University of Texas at Arlington,
Arlington, Texas 76019, USA\\
$^3$Pacific  Northwest  National  Laboratory,  Richland,  WA  99352,  USA\\
$^4$Fermi National Accelerator Laboratory, Batavia, IL 60510}

\medskip

 \begin{Abstract}
\noindent We introduce a novel particle detection concept for large-volume, fine granularity particle detection: The Ion Fluorescence Chamber (IFC).  In electronegative gases such as SF$_6$ and SeF$_6$, ionizing particles create ensembles of positive and negative ions.  In the IFC, positive ions are drifted to a chemically active cathode where they react with a custom organic turn-on fluorescent monolayer encoding a long-lived 2D image.   The negative ions are sensed electrically with course resolution at the anode,  inducing an optical microscope to travel to and scan the corresponding cathode location for the fluorescent image. This concept builds on technologies developed for barium tagging in neutrinoless double beta decay, combining the ultra-fine imaging capabilities of an emulsion detector with the monolithic sensing of a time projection chamber. The result is a high precision imaging detector over arbitrarily large volumes without the challenges of ballooning channel count or system complexity. After outlining the concept, we discuss R\&D to be undertaken to demonstrate it,  and explore application to both directional dark matter searches in SF$_6$ and searches for  neutrinoless double beta decay  in large $^{82}$SeF$_6$ chambers. 
\end{Abstract}

\bigskip

\def\thefootnote{\fnsymbol{footnote}}
\setcounter{footnote}{0}

\section{Introduction}

Time Projection Chamber (TPC)~\cite{fancher1979performance} detectors have become pervasive for precision particle detection applications.  One of the great virtues of the TPC that has contributed to its ubiquity is the fact that a large 3D volume may be imaged from a single plane of sensors.  Electrons drifting to this plane carry information about the trajectories of ionizing particles that traverse the gas, and the image may be reconstructed from an instrumented anode plane no matter where in the detector volume the original event occurred.    This allows for favorable scaling behaviour of detectors, enabling viable systems up to even the kiloton scale~\cite{acciarri2016long}.

This method of extracting information does not come without some associated technical overhead.  In TPC detectors with long drift lengths, maintaining the image quality over a large drift distance requires purification of the gas to part-per-billion levels in electronegative contaminants to avoid electron attachment losses.  Over long drift distances, diffusion will  compromise image sharpness in both longitudinal and transverse directions.  And sensitive electronics are needed that digitize each channel with microsecond resolution and few hundred electron equivalent noise charge per sample, which can be costly and complex for large devices.  

None of these technical requirements have proven to be particularly limiting in the majority of TPC use cases, and as such the TPC is pervasive as the go-to detector for precision event imaging at large scale today. However, there are some scenarios where future desirable experiments may push the TPC concept toward the limits of credibility. Two notable examples are the searches for directional recoils of weakly interacting massive particle (WIMP) dark matter~\cite{vahsen2021directional} and  gigantic kiloton-scale TPCs for neutrinoless double beta decay~\cite{avasthi2021kiloton}.  In this white paper we consider what makes these particular applications inelegant for conventional TPCs, and propose a novel detection concept that aims to overcome this possible inelegance.

The search for directional recoils induced by dark matter particles scattering on light gases has long been recognized as a promising methodology for dark matter searches, and several ideas exist for realizing this concept~\cite{recoilwhite}.  Reconstruction of directional recoils is one of few techniques that allows dark matter detectors to penetrate below the sensitivity floor imposed by an ever-present background of solar neutrinos~\cite{boehm2019high}. However, realizing a sufficiently sensitive directional dark matter detector at the required scale is known to be extremely technically challenging.  The difficulty of such experiments stems, in part, from the very low energy of the recoil. These recoils range from $\mathcal{O}(1) - \mathcal{O}(100)$ keV over most plausibly experimentally accessible WIMP parameter space.  In order to observe directionality of this recoiling nucleus, it must travel far enough in the gas before coming to a stop through ionizing energy losses that the shape of the track is not destroyed by diffusion before charge reaches the detection plane.  To minimize the severity of diffusion, electronegative gases have been used as carriers in place of electrons~\cite{martoff2005negative}.  Since ions drift thermally, due to their high mass and collision cross sections, ion drift significantly reduces the scale of diffusion over a fixed drift distance.  At reduced electric fields up to at least 1000~kV/cm bar, drifting ions will be in thermal equilibrium with the bulk gas and so the scale of diffusion will be close to the thermal limit:

\begin{equation}
    \sigma_D=\sqrt{\frac{4 k_B T L}{e E}}
\end{equation}

Negative ion TPCs are typically instrumented with pixelized readouts that have a pitch on the 100$ \mu$m scale~\cite{ikeda2018study,ligtenberg2021properties}.  The requirement that a recoil with tens-to-hundred  keV travel a significant fraction of a millimeter in order to be resolved implies that gas pressures must be at or below 100 Torr. Coupling this operating parameter with the now extremely low limits on WIMP nucleus cross sections of order 10$^{-44} $cm$^2$ (provided by ton-scale noble liquid detectors and solid state techniques~\cite{roszkowski2018wimp}) implies that detectors of vast physical size will be needed to probe the interesting parameter space.   What is problematic about this scenario is the combination of the requirements of this huge detector size with ultra-fine resolution.  While neither seems impossible on its own, their combination leads to a ballooning electronic channel count and hence electronic complexity that appears very difficult to realistically achieve.

An entirely different physics application targeting rare events at a different energy scale is the realization of ultra-large searches of neutrinoless double beta decay that would be needed to probe half-lives at the 10$^{30} $yr level~\cite{avasthi2021kiloton}. This is required if the field intends to achieve sensitivities to normal mass ordering part of parameter space under the light Majorana neutrino exchange mechanism~\cite{jones2021physics}. Accepting that acquisition of enough isotope for such an experiment will be a major challenge to be overcome, we may ask: what else would technically stand in the way of building such a large time projection chamber?  For these experiments, the amount of energy available per event is larger than the example of directional dark matter searches: 2.4 MeV in xenon, for example, with similar MeV-energy scales in most other isotopes of interest.  The primary difficulties in construction of very large neutrinoless double beta decay experiments lie in:
\begin{itemize}
    \item Suppressing radiogenic backgrounds to the vanishingly small levels needed for sensitivity through either active (topological identification~\cite{ferrario2016first}, energy resolution~\cite{alvarez2013near}, barium tagging~\cite{mcdonald2018demonstration,Chambers:2018srx}) or passive (self-shielding ~\cite{adhikari2021nexo}, radiopurity campaigns~\cite{alvarez2013radiopurity,kharusi2018nexo}) means; 
    \item Achieving sufficient energy resolution over the full volume to negate the background from two neutrino double beta decay events - in practice better than around 2\% FWHM at the ton-scale, but better still as detectors become larger; 
    \item Ensuring no other background sources creep into the experiment, accounting for the above issues. 
\end{itemize}

Measuring energy with 2\% FWHM resolution implies that collection of charges on an electrically conducting anode alone will not suffice.  In liquid xenon TPCs, recombination fluctuations mean that charge and light must both be collected and anti-correlated~\cite{anton2020measurement}, and even here the energy resolution has not been shown to be better than 3\% FWHM~\cite{albert2018search}. In gaseous xenon TPCs at up to 10 bar, absence of recombination means that collection of charge alone is sufficient to provide energy resolutions at the sub-1\% FWHM level~\cite{renner2019energy}. However the resolution required for such a precise measurement of the charge is unattainable in most common readout schemes. Both a readout on sense wires followed by electronic amplification, and avalanche gas gain followed by charge collection, introduce too large a fluctuation to achieve the needed resolution.  Electroluminescent gain is typically used, and this in itself seems technically challenging to realize at vast detector scales, both in terms of construction of an extremely large electroluminescent region, and in manufacturing a physically large and high density 175~nm-sensitive pixelized plane to monitor it with. The application of kTon-scale TPCs for this application is therefore challenged by a similar practical problem to the directional dark matter recoil case, albeit four orders of magnitude higher in energy - the need to sensitize a very large surface area with fine pitch.

One option to overcome such problems is by a brute force: or in other words, to rely on continuing industry-driven improvements in electronic fabrication to solve the channel count problem. An experiment need only to spend the money required for the millions of readout channels needed to instrument the system at the desired resolution,  noise level, and low radiogenic background. It is notable that even if possible this is likely a somewhat wasteful solution, since in the case of a well constructed dark matter or double beta decay experiment most of the channels will be doing nothing most of the time; they just need to be there, fully active, and waiting for an event to happen. 

This paper considers a different approach, which involves implementing  a new type of detector with ultra-fine imaging resolution, with imaging is delayed until after the event location has been identified with a higher speed but lower precision sensor. This may allow realization of micron-scale imaging but with a sparsity of sensors, in order to cover a much larger area with the precision needed to make ultra-large directional dark matter and normal-mass-ordering scale neutrinoless double beta decay experiments. This concept, which in some sense may be thought of as a marriage between some of the best qualities of TPC detectors~\cite{gonzalez2018gaseous} with those of emulsion detectors~\cite{alexandrov2021directionality}, is called the Ion Fluorescence Chamber (IFC) which we now describe.

\section{Detection concept}

\begin{figure}[t]
\begin{center}
\label{Lego}
\includegraphics[width=0.99\columnwidth]{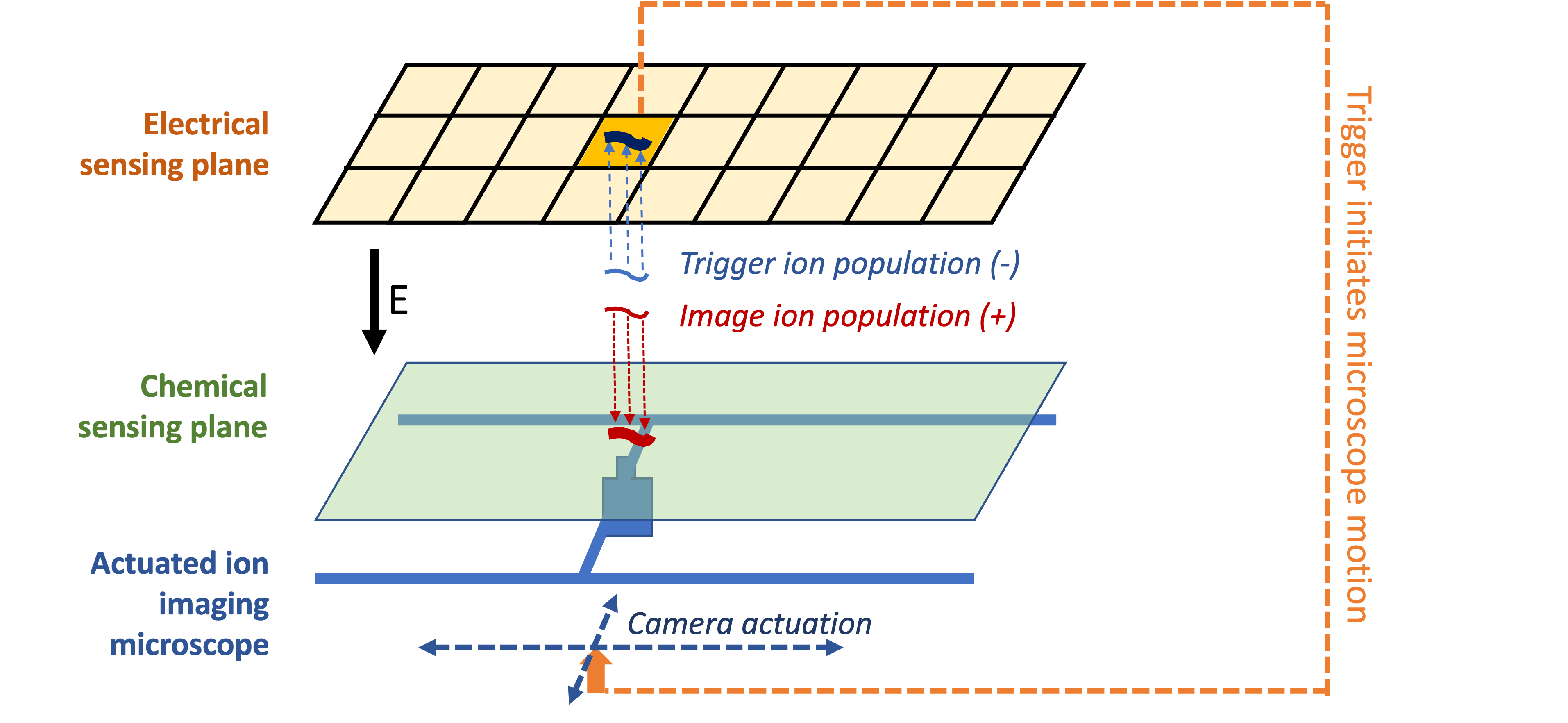} 

\caption{Basic concept of the Ion Fluorescence Chamber (IFC).  The trigger ion population is detected electrically with coarse resolution and in real time; this initiates actuation of a sensor that provides a delayed microscopic image of the image population through turn-on fluorescence sensing.\label{fig:BasicConcept}}
\end{center}
\end{figure}

A conceptual picture of the proposed detection method is shown in Fig.~\ref{fig:BasicConcept}.  The system is a TPC where both the positive and negative ion signals from interactions are collected, through two distinct sensing techniques.  Which role is served by the positive vs negative ions will depend on what proves to be the optimal chemical configuration for the ion sensing layer.  Conceptually one can imagine a detector operating in either negative-trigger or positive-trigger mode, but for concreteness we will consider negative-trigger mode in this work.  

In an interaction in an electronegative gas such as SeF$_6$ or SF$_6$, electrons and positive ions are produced through initial ionization, with the electrons then quickly capturing on the electronegative molecules to yield populations of XF$_N^+$ and XF$_N^-$ ions, where in general the distribution between different $N$ may be a rather complex one~\cite{nygren2018neutrinoless,Phan}.  The positive and negative ions will drift in opposite directions, to two planes with distinct function.

The {\bf trigger population} of ions is collected by a low spatial resolution electronically digitized plane of readout electrodes that provides event localization at a ~5 cm scale.  The low spatial resolution implies only a modest readout channel count is required, even for a very large detector.  While the trigger population could be either the positive or negative species, there is a notable advantage to using the negative ions for this purpose.  This is because negative ions can be stripped in a high field region to produce avalanche gain to amplify the electrical signal from the small ion population.  For the double beta decay application this is less of a concern, and either positive or negative ions can likely be sensed electronically with no gas gain, but for dark matter searches, observing the small electronic signal over noise from a large capacitance pad suggests stripping and gain will be a significant benefit.   In either case, the trigger provides two pieces of information: 1) fiducialization of the event in X and Y via transverse position, and in Z via measurement of longitudinal diffusion in gas (a technique demonstrated in xenon gas for 45 keV $^{83m}$Kr X-Rays~\cite{alvarez2013near}); 2) approximate measurement of the transverse position which, following some loose hardware-level trigger cuts, initiates motion of the actuated camera to realize precision imaging on the opposite plane.

The {\bf image population} is detected at the opposite detector plane via a fluorescent ion-sensing layer.  This layer, which fully covers the cathode, is comprised of molecules that are non-fluorescent but become fluorescent upon chemical reaction with the ions of interest.  The ions arriving at the cathode can then be imaged with resolution at the sub-micron level to the point that, in principle, every individual arriving ion can be counted.  If turn-on fluorescence efficiency upon ion arrival is sufficiently high this will provide both a precise energy measurement and a topological image limited only by diffusion.  Since ions drift with far less diffusion than electrons, they carry a more pristine representation of the spatial form of the event than achievable with an electron TPC, enabling precise directionality and $dQ/dx$ measurement within the 2D plane.   Moreover, because the ion image is long lived at the cathode and remains in place until the the switched-on molecules are photo-bleached by laser excitation~\cite{mcdonald2018demonstration}, a very large plane can be read out with only a small number of scanning systems.  This long retention time of the ion image is, in effect, what enables the detector to be simultaneously very large and very precise, overcoming the limitations of traditional TPC readout methodologies in this regard.

\section{Chemistry of the ion sensing layer}

\begin{figure}[t]
\begin{center}
\label{Lego}
\includegraphics[width=0.95\columnwidth]{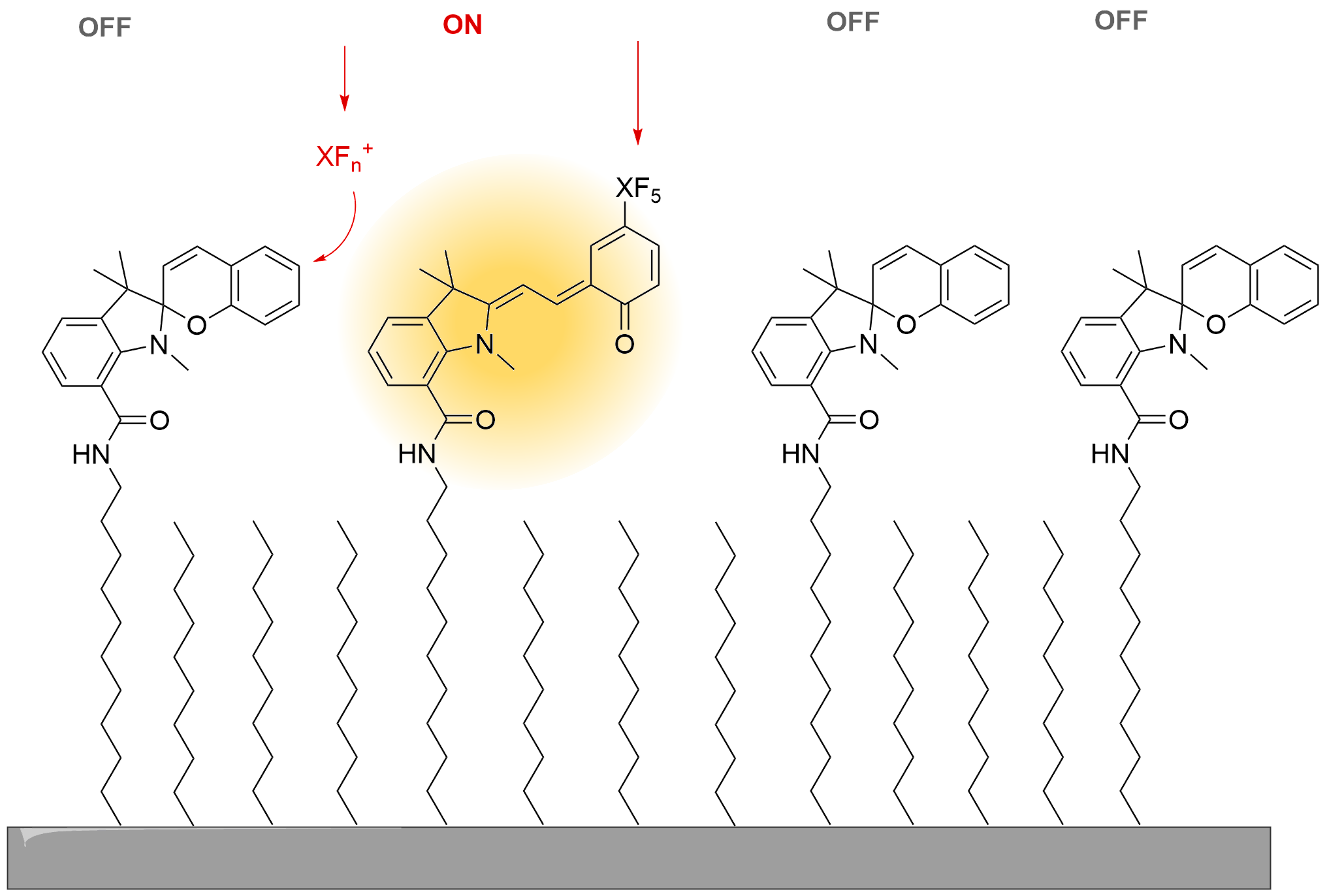} 

\caption{The supramolecular organic monolayer that is used to make the fluorescent image of the arriving ions at the detection plane. In this case we imagine a class of sensors constructed for sensing the positive ions produced in XF$_N$ where X=S or Se, motivated by dark matter searches and neutrinoless double beta decay searches, respectively. \label{fig:monolayer}}
\end{center}
\end{figure}

The ion sensing layer is a conceptual cousin and technological derivative of the barium tagging monolayers under development for neutrinoless double beta decay searches by some of us~\cite{byrnes2020demonstration,thapa2021demonstration,thapa2019barium}, but with a number of important differences.  For that application, single ion detection at scanning surfaces has been demonstrated over mm$^2$ areas with as fine as 2~nm spatial resolution, using custom-designed fluorophores with dry fluorescent response to target metal dications~\cite{thapa2019barium}.   Variants that change color, rather than switching from ``off'' to ``on'' upon binding, have also been explored by others~\cite{rivilla2020fluorescent}.  For the present application, ion sensing molecules must be developed that transition from non-fluorescent to fluorescent upon binding with the host gas ions, rather than metal cations. The absence of reactions with the neutral gas molecules, long sensor lifetime, low intrinsic background, and at least moderately high efficiently for ion capture and fluorescence are the key driving criteria.  Chemical specificity between different positive ions which is an important quality for the barium tagging application~\cite{thapa2021demonstration}, is not a strong driver here - and it may even be a limitation in gases that generate multiple different positive or negative ion populations.  A great number of possibilities exist for sensing either positive or negative ions in a wide variety of molecular gases.  In the interest of studying a concrete system, we will concentrate this first discussion on two gases that are chemically similar and have been previously proposed for dark matter searches and neutrinoless double beta decay searches respectively, SF$_6$\cite{Phan} and SeF$_6$ that has been enriched in the isotope $^{82}$Se.

There are several possible avenues that appear promising for sensing XF$_N^+$ and XF$_N^-$, N$\leq$6.  In all cases the required supramolecular structure is similar: a self-assembled organic monolayer of sensor molecules on a thin substrate for fluorescence imaging.  Such layers can be grown though either vapor deposition or from solution on a flat transparent substrate such as quartz or glass.  Here we focus on one plausible method of XF$_N^+$ cation sensing using an organic molecular monolayer of the type shown in Fig.~\ref{fig:monolayer}.   The monolayer is formed from a series of sensing sites separated by passive spacer groups. The required surface density of sensors should be large enough that each SeF$_N^+$ cation has a high probability of binding, yet not so large that the sensing groups interfere with one another through collective effects.  Incorporation of ion-mobile surface elements may enable further reductions in surface density of sense molecules.

\begin{figure}[t]
\begin{center}
\label{Lego}
\includegraphics[width=0.99\columnwidth]{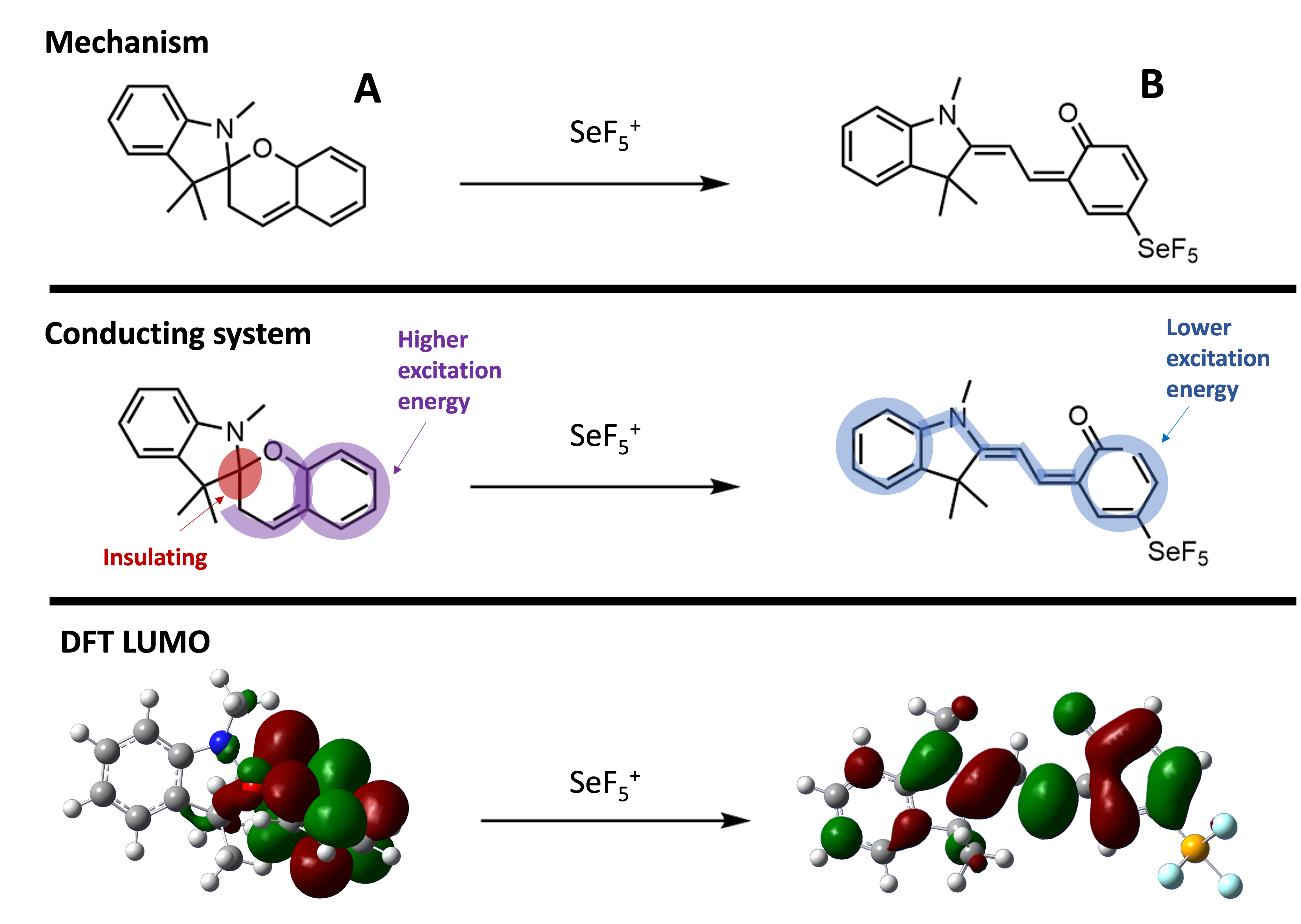} 

\caption{Molecular design concept for the XF$_N^+$ chemosensor.  Reaction with the ion of interest stabilizes the open configuration B of the molecule, which has a larger degree of HOMO/LUMO delocalization than the unbound molecule A. This enables turn-on fluorescence sensing upon ion arrival when probed with a visible excitation laser. \label{fig:molecule}}
\end{center}
\end{figure}

XF$_N^+$ cations have special reactivity with Lewis bases and electron-rich aromatic molecules in the gas phase.\cite{stone1989high} Upon binding to a sensing site, XF$_N$ substituents--whether they are unchanged or after eliminating small molecules--remain strongly electron-deficient and are predicted to cause a shift in electron density within the molecule \cite{gilbert2021electron}.  In the case of the proposed sensing molecule, such an electronic shift should induce a transition from a non-fluorescent ({\bf A}) to a fluorescent ({\bf B}) configuration \cite{winkler1998photodynamic}. Fig.~\ref{fig:molecule} illustrates in a schematic way the reaction-activated mechanism for turn-on fluorescent behaviour in this system.  The withdrawal of electron density from the aromatic system induces a change in the bonding configurations around the central carbon atom of the molecule that transitions the central region of the molecule from an effectively insulating configuration to an effectively conducting one. This creates a substantial difference in the delocalization scale of the frontier electrons orbitals, resulting in excitation by longer wavelength visible light. Fluorescence is possible when an un-quenched electron excitation is possible between two orbitals, usually the highest occupied molecular orbital (HOMO) and lowest unoccupied molecular orbital (LUMO). A high efficiency for transition requires significant spatial overlap between these orbitals. In cases where there is large spatial overlap between one orbital below the HOMO (HOMO-1) and the LUMO, excitation is possible but the resulting emission is likely quenched by the higher potential energy HOMO electrons.

The shapes of these orbitals determine the expected spectral characteristics.  Speaking loosely, the size of the ``box'' containing the electron responsible for fluorescence determines the absorption and emission wavelengths. When multiple transitions are possible, there may be multiple spectral peaks.  In the system we have considered here, this box becomes much larger when the molecule is bound to the target ion.  This is shown in cartoon form in Fig.~\ref{fig:molecule} middle, and verified with density functional theory calculations in the lower panel.   The computed LUMO with density functional theory (Fig.~\ref{fig:molecule} bottom), which is the orbital that would receive the promoted electrons upon fluorescence excitation, becomes dramatically enlarged upon binding to XF$_N$.

This change in the localization scale of the electron within the molecule is matched by a corresponding change in the energy scale for fluorescent excitation. Whereas the unbound molecule is expected to fluoresce in the deep UV, the bound one will fluoresce at the visible wavelengths ($\sim$570 nm excitation and $\sim$610 nm emission). When probing with a visible laser, as is the natural choice for fluorescence sensing, the effect will be a fluorescent switch-on effect upon reaction with positive ions. 

\begin{figure}[t]
\begin{center}
\label{Lego}
\includegraphics[width=0.99\columnwidth]{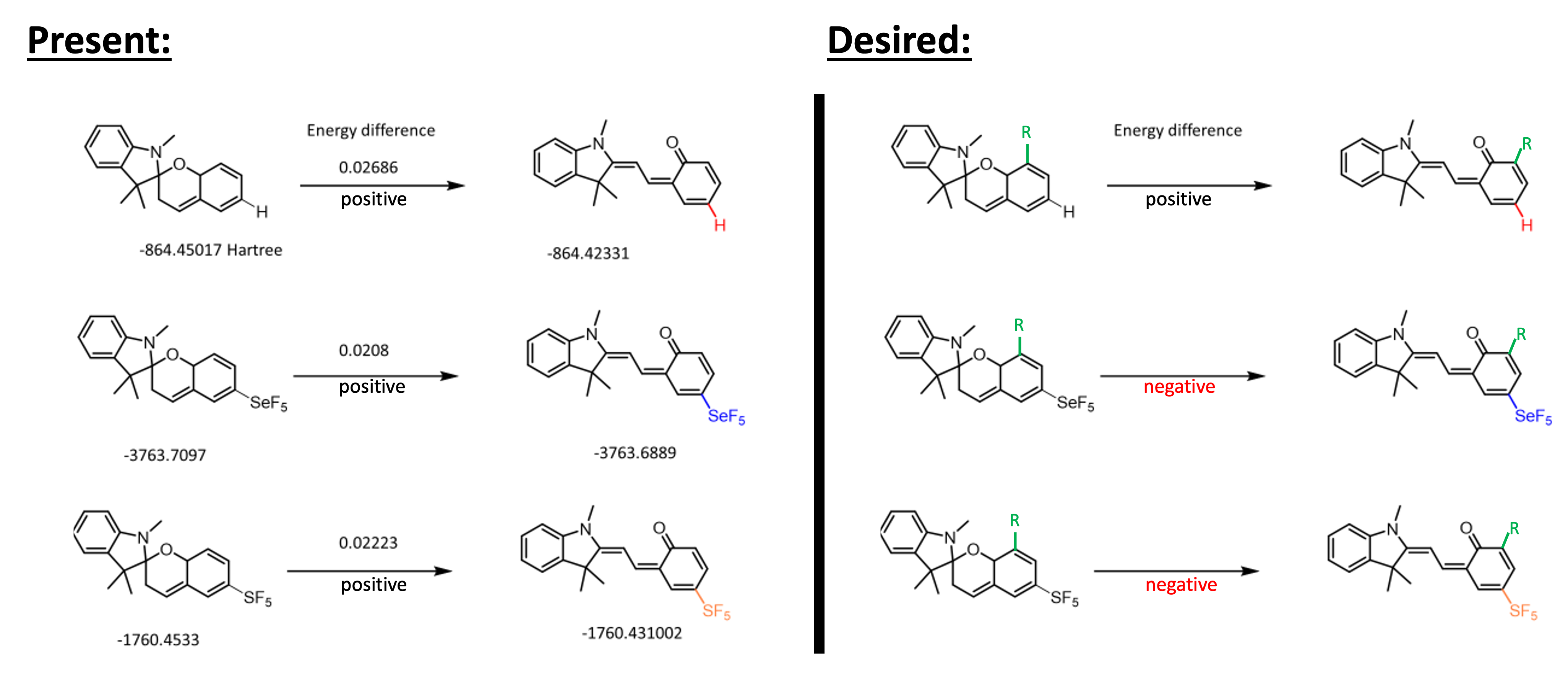} 

\caption{Illustration of the molecular design problem that remains to be solved with this sensor, left. While the open, fluorescent configuration is indeed stabilized upon binding as is desired, the closed configuration remains the ground state.  This behaviour can be tuned by adding electron-withdrawing groups $R$ to adjust relative stability of the two states until strong turn-on fluorescence response is achieved. This computational work is ongoing. \label{fig:positivedesired}}
\end{center}
\end{figure}

There are other properties this molecule should exhibit if it is to be suitable for this purpose, in addition to the enabling feature, an increasing LUMO delocalization between {\bf A} and {\bf B}. Achieving all of these requirements in one molecule will require extended study, both theoretically and chemically.  One feature that is imperfect about the molecule shown in Fig.~\ref{fig:molecule} is that while the open state is stabilized by XF$_N^+$ addition by a significant fraction of an electronvolt, the ground state of both XF$_N^+$ bound and unbound forms is the closed configuration, separated from the excited open state by around 0.54~eV.  Modifications must be made to the molecular design to maintain the stabilization of the open form by XF$_N$ binding, while ensuring the closed form is preferred as the ground state for the unbound system. Prior efforts to achieve metal ion conversion of {\bf A} to {\bf B} was achieved by incorporation of nitro-substituents~\cite{feuerstein2019investigating}.  This is shown schematically in Fig.~\ref{fig:positivedesired}-where may equal R=NO$_2$-and is the subject of ongoing computational explorations.  

\begin{figure}[t]
    \centering
    \includegraphics[width=0.7\columnwidth]{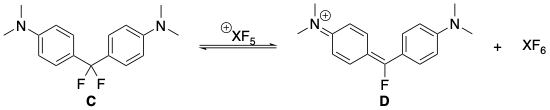}
    \caption{Alternate XF$_N^+$ reactive chemosensor exploration. Reaction of the cation with one halogen (fluorine shown) within the central carbon of \textbf{C} is predicted to result in dehalogenation. The loss of a fluorine atom creates a highly fluorescent cation, \textbf{D} under visible light.}
    \label{fig:F2Michlerketone}
\end{figure}

A related XF$_N^+$ reaction-based fluorophore class under investigation is represented by the halogenated Michler ketone, \textbf{C}, and related diaryl and triaryl cationic sensing agents \textbf{C}, shown in Fig.~\ref{fig:F2Michlerketone}. This system is suggested by ion collision studies between gas-phase halogens and SF$_5^+$ \cite{howle2005selected}. When sensing agent \textbf{C} is dehalogenated by gas-phase binding with XF$_N^+$ it should result in a neutral XF$_6$ molecule and cationic fluorescent version of the sensing agent, \textbf{D}.  Here, the carbon bearing two fluorine atoms acts as an insulating group separating two aniline rings that absorb in the UV region. Upon loss of the fluorine, the insulating structure is removed and the entire molecule becomes a visible-light fluorophore that absorbs ~460 nm and provides a single emission peak ~550 nm.\cite{geng2020preparation} These XF$_N^+$ triggered fluorophores are similarly being pursued by a suite of computation, organic synthesis, and photochemical experiments.

\section{Optical considerations}

Once a suitable molecular monolayer exhibiting the required properties in gas can be realized, a system must be implemented to image this layer over very large detector scales.  The requirements for the optical system appear demanding but not insurmountable.  Actuation of the microscope incorporating both excitation and detection elements over the full detection plane is required.  The objective must be able to be located with a precision smaller than the size of the field of view in the XY direction to ensure a reliable scan of the area aligned with the trigger region; and with a precision similar to the depth of focus in the Z direction to ensure focus on individual emitters can be achieved.

Experience with development of in-gas optical systems for barium tagging~\cite{byrnes2020demonstration} suggests that most likely this will be achieved with a two-stage focusing system, first a rough placement with $\sim$~100~$\mu m$ precision using a mechanically geared system, followed by a second and more finely controlled placement using piezoelectric stages.  Under this scheme, after moving to approximately the right location facing the activated electronic element on the trigger plane, a scan will need to be made over the focal plane to collect a tessellated image large enough to cover the event region.  A sparse distribution of point-like calibration standard candles such as fluorescent microspheres may assist in quickly achieving focus, without substantially obscuring the image. Assuming the use of high magnification objectives with high numerical aperture that would be needed for the case of imaging every ion individually, the field of view may be around 100~$\mu m$ wide in the transverse directions and 1~$\mu m$ in depth.  It is also unlikely to be flat in $z$ across the transverse imaging region.  Placement of the objective using the rough mechanical stages is unlikely to be repeatable with micron-scale precision. However, after this rough placement, the focal plane at each XY can be obtained using interpolated auto-focus techniques of the kind that have already been realized in the context of barium tagging R\&D.  

\begin{figure}[t]
\begin{center}
\label{Lego}
\includegraphics[width=0.7\columnwidth]{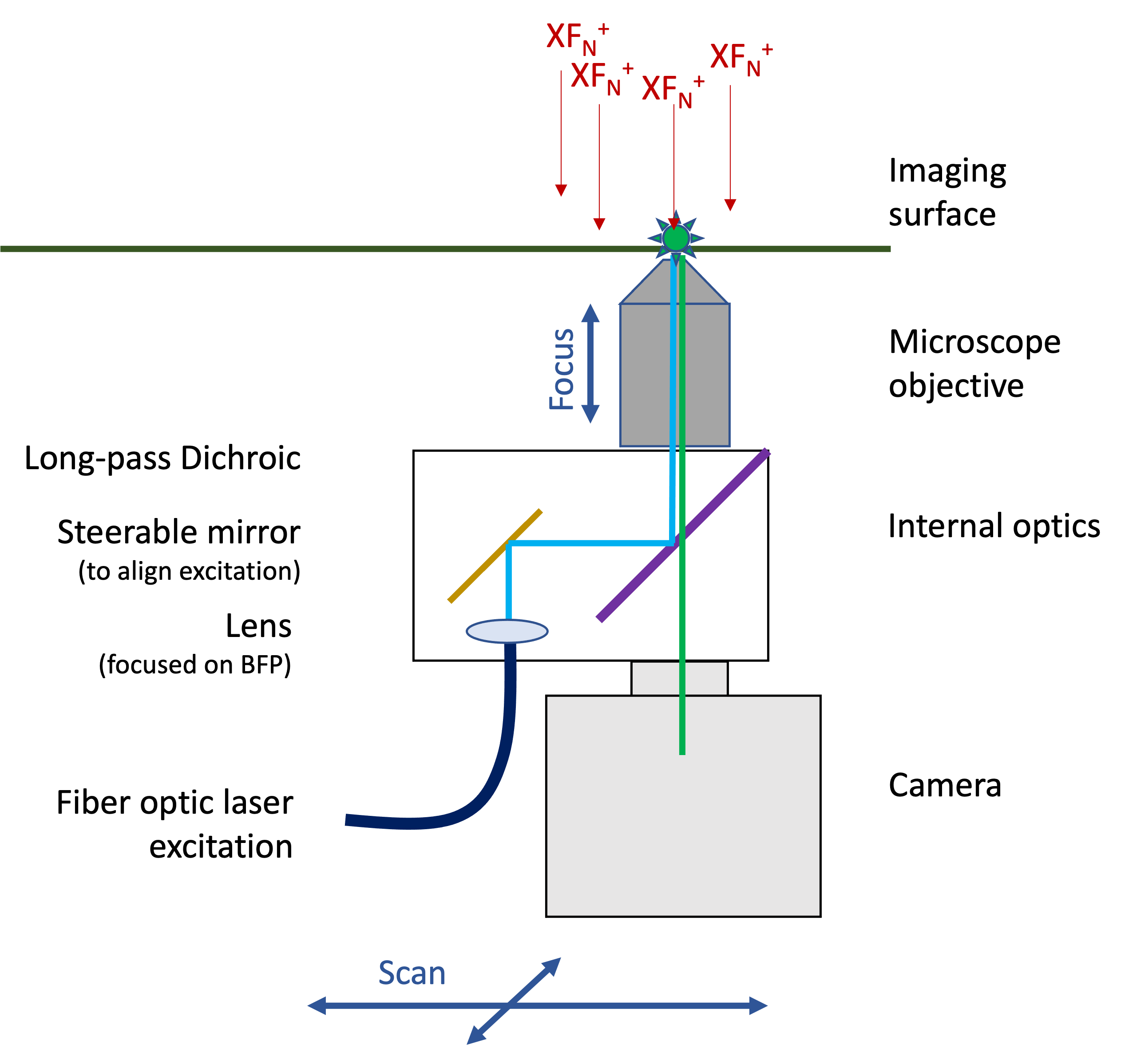} 

\caption{Example of an optical system that may be used for imaging the fluorescent monolayer. This system is rather similar to the system we have proposed and developed for barium ion tagging in high pressure xenon gas.\label{fig:Optics}}
\end{center}
\end{figure}

A sketch of the optical system that may achieve single ion imaging precision after rough positioning is shown in Fig~\ref{fig:Optics}.  In this system, laser excitation light is delivered via single-mode optical fiber and positioned by a piezeoelectrically steerable mirror onto a short-pass dichroic mirror. The short-wavelength excitation light reflects off the dichroic mirror and is focused on back-focal plane (BFP) of a high numerical aperture 100X objective. The objective is positioned in the Z direction using a piezoelectric stage to achieve focus on the imaging plane, though notably the depth of focus of the excitation lens is long enough that uniform illumination of the sample by parallel rays is expected to be maintained.  Total internal reflection fluorescence (TIRF) imaging is also a possibility with this system by focusing off-center of the BFP of the objective.

A high quantum efficiency camera such as an EMCCD will be mounted on the XY-positionable stage that will actuate behind the transparent imaging plane. It is likely that this camera will need to be enclosed within a hermetically sealed volume that moves within the detector vessel, in order to both to avoid poisoning the gas purity and compromising the operating life of the camera by interaction with the gas.  This does not appear to be particularly difficult to achieve, especially since the light being detected by this camera is in the visible wavelength range, though notable issues such as removal of the heat load produced by cooling of the low-noise EMCCD element will need to be managed.  Our experience developing in-gas microscopy systems suggests these issues can all be addressed with careful R\&D, to yield a precision-positionable XY scanning optical system.

\begin{figure}[t]
\begin{center}
\label{Lego}
\includegraphics[width=0.99\columnwidth]{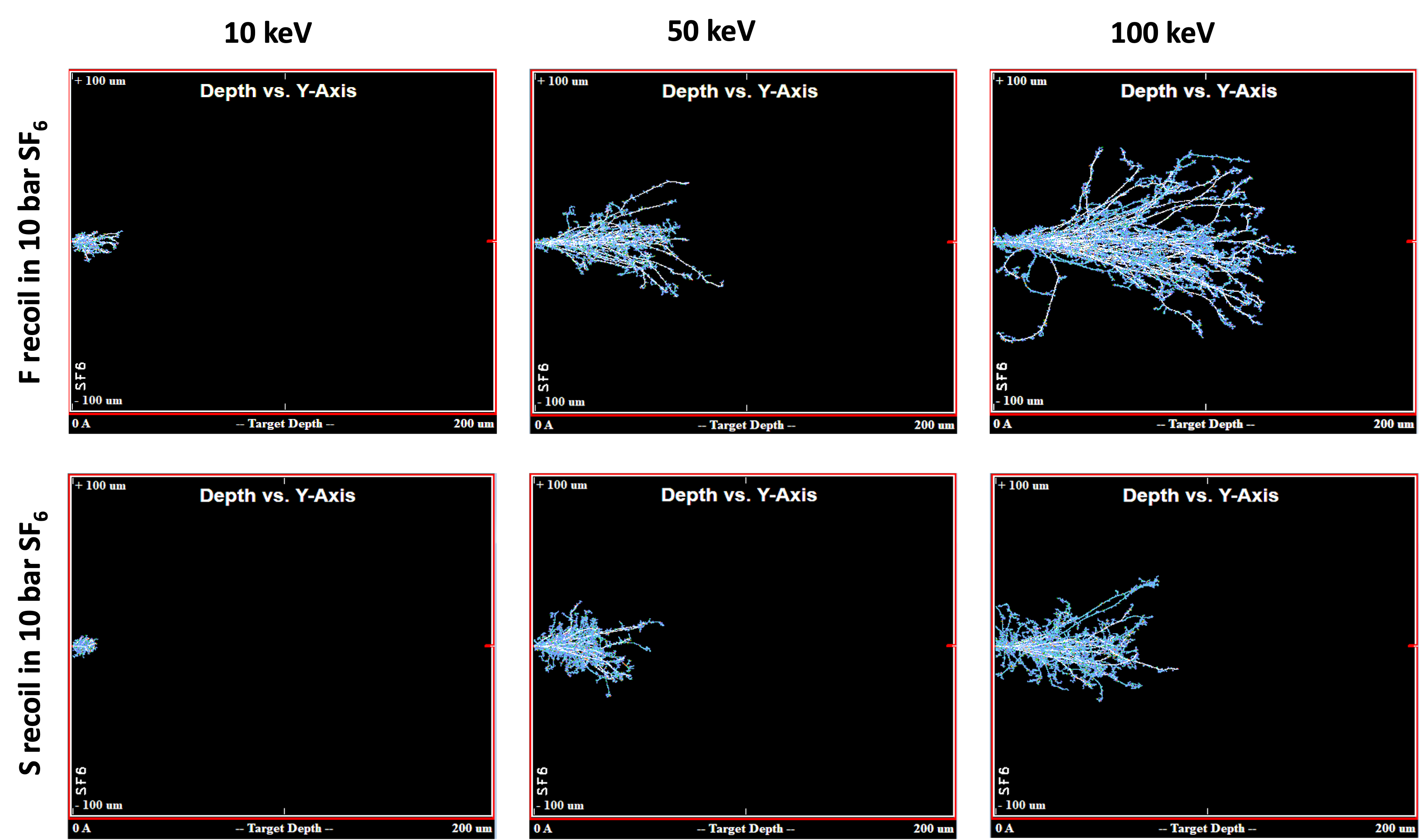} 

\caption{The Stopping and Range of Ions in Matter (SRIM) simulations of nuclear recoils in 10 bar SF$_6$ gas.  The panels show recoil trajectories in an ensemble of 100 events simulated at 10, 50 and 100 keV (left, middle, right) with S and F recoils (top, bottom) respectively.  Even at 50~keV, significant directionality is observed. \label{fig:SRIMSim}}
\end{center}
\end{figure}

\section{Performance explorations}

In this section we offer a rather ``fast-and-loose'' discussion of possible performance estimates for the dark matter and double beta decay applications.

\subsection{Dark matter recoil detection with SF$_6$}

The gas SF$_6$ has been identified as a highly promising gas for negative ion dark matter TPCs and studied extensively in Ref.~\cite{Phan}.  Our concept is essentially a novel method for reading out the type detector described at some length in that work, but with a greater sparsity of sensors and intrinsic position resolution. Many of the detailed considerations about SF$_6$ detectors elucidated there will apply in this case as well.

Dark matter recoils in SF$_6$ may be recoils of either the S nucleus or the F nucleus.  The higher the energy of the recoil, the stronger the directionality of the signal.  At higher gas pressures, the events are smaller in physical size, but in principle recoil directionality is maintained. For example, Fig.~\ref{fig:SRIMSim} shows The Stopping and Range of Ions in Matter (SRIM) simulations of nuclear recoils in 10~bar SF$_6$ gas. It is clear that even though the events are rather compact  (tens of microns), some directionality of the recoil track is present even at 50~keV.  

\begin{figure}[t]
\begin{center}
\label{Lego}
\includegraphics[width=0.6\columnwidth]{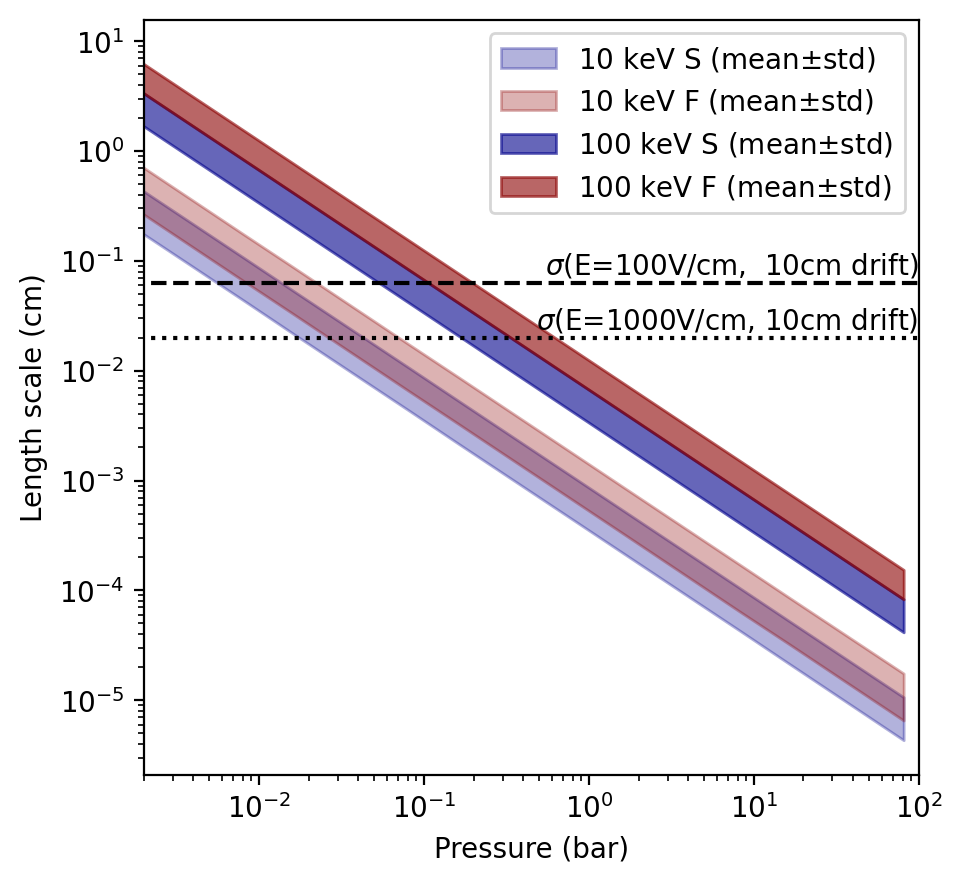} 

\caption{Comparison of the relevant length scales of recoil distance-of-travel (bands) and the distance scale associated with thermal diffusion over 10~cm (black lines), for 10 keV (light) and 100 keV (dark) nuclear recoils. The crossing point of these two distance scales suggests the largest plausible operating pressure for a specific drift distance, with direct implications for total detector volume.\label{fig:RecoilLengths}}
\end{center}
\end{figure}

While microscopic reconstruction of fluor distributions with length-scales of microns is clearly viable with existing technology, the key question is whether diffusion can be controlled to a level that any directionality present is not entirely erased during drift. Since the thermal diffusion limit does not depend on pressure whereas track length does, in practice this implies for a given recoil energy scale, a maximal pressure and hence minimal event size at which the detector can run and maintain directional sensitivity.  Fig.~\ref{fig:RecoilLengths} compares the thermal diffusion limits at drift fields 100~V/cm and 1000~V/cm with the expected ranges of 10~keV and 100~keV  nuclear recoils in SF$_6$ simulated in the SRIM software package~\cite{ziegler2010srim}.  The band in each case shows mean$\pm$standard deviation.  We see that in practice, pressures less than 0.1 bar are likely required for practical application.  This low pressure implies a physically large detector; however, with this readout scheme it is one that does not have to have a completely sensitized anode or unrealistically large channel count, so implementation at the required scale appears plausible. 

It is beyond the scope of this work to make a full analysis of the expected directional sensitivity of this device, though the general principles this would require are similar to those that have been expanded upon at length by others \cite{Mayet:2016zxu}.

\subsection{Neutrinoless double beta decay searches with SeF$_6$}

The possibility of using negative ion TPCs for neutrinoless double beta decay was outlined in Ref.~\cite{nygren2007negative}, and the prospects for neutrinoless double beta decay with SeF$_6$ as a working medium in particular were explored in Ref.~\cite{nygren2018neutrinoless}.  Here we summarize the key conclusions of that study, and emphasize the relevance of the detector concept outlined in this paper for realizing the promise of this working medium.

The intrinsic energy resolution in ionization was evaluated and found to be 0.34\% FWHM, to be compared with xenon gas at 0.28\% FWHM. This assumes that fluctuations from recombination are not limiting, which is estimated there to be a contribution of only $\sim$0.05\% FWHM at 10 bar of gas pressure.  Thus, as in high pressure xenon gas, instrumental effects rather than gas microphysics are likely to determine the ultimate energy resolution of the experiment.

Accounting for production of the various SeF$_N$ anions and cations, each species with a different specific ionization energy, the number of ions produced at the Q value for double beta decay in $^{82}$Se is estimated to be around N$\sim$91,000, with the above energy resolution already accounting for fluctuations in energy partitioning between different species.  If the ion collection and counting efficiency is $\epsilon$, the Poissonian fluctuations in energy resolution will be $E_{FWHM}\sim 2.355\sqrt{\epsilon N}/\epsilon N$. A 50\% ion collection efficiency thus implies approximately a 1\% FWHM energy resolution, comfortably within the requirements of a large-scale neutrinoless double beta decay search.  Much smaller ion imaging efficiencies than this would likely be problematic for rejecting the two neutrino double beta decay background. A 50\% per-ion imaging efficiency can thus be take to be a (very approximate) target performance metric for viability of this approach as a neutrinoless double beta decay search.

Topological imaging in an $^{82}$SeF$_6$ gas TPC was also explored in Ref.~\cite{nygren2018neutrinoless}.  At 10~bar, the electrons emitted in $^{82}$Se double beta decay travel slightly further than those emitted by $^{136}$Xe due to the higher Q-value of the decay.  This leads to a somewhat improved topological event identification performance than in $^{136}$Xe gas for equivalent background acceptance, given perfect resolution on initially deposited ionization charge.  

Two factors beyond the increased Q-value promise to significantly improve the topological event identification quality beyond that of $^{136}$Xe gas, however. First, diffusion of positive and negative ions is purely thermal, and much smaller than the diffusion of electrons in xenon gas.  Although much of the effect of diffusion can be removed by modern deconvolution methods~\cite{simon2021boosting}, there is no doubt that a dramatic reduction in diffusion coupled with dramatic enhancement in position resolution at the detection plane should lead to a substantial increase in power to topologically identify events.  Second, the far lower levels of background in an  $^{82}$SeF$_6$ double beta decay search due to the Q-value being far above most radiogenic gamma ray lines (especially those associated with $^{214}$Bi and $^{208}$Tl) implies that topological cuts for signal selection can likely be much looser.  These factors together promise a significant increase in sensitivity for similar exposure, relative to TPCs using $^{136}$Xe gas.

\begin{figure}[t]
\begin{center}
\label{Lego}
\includegraphics[width=0.8\columnwidth]{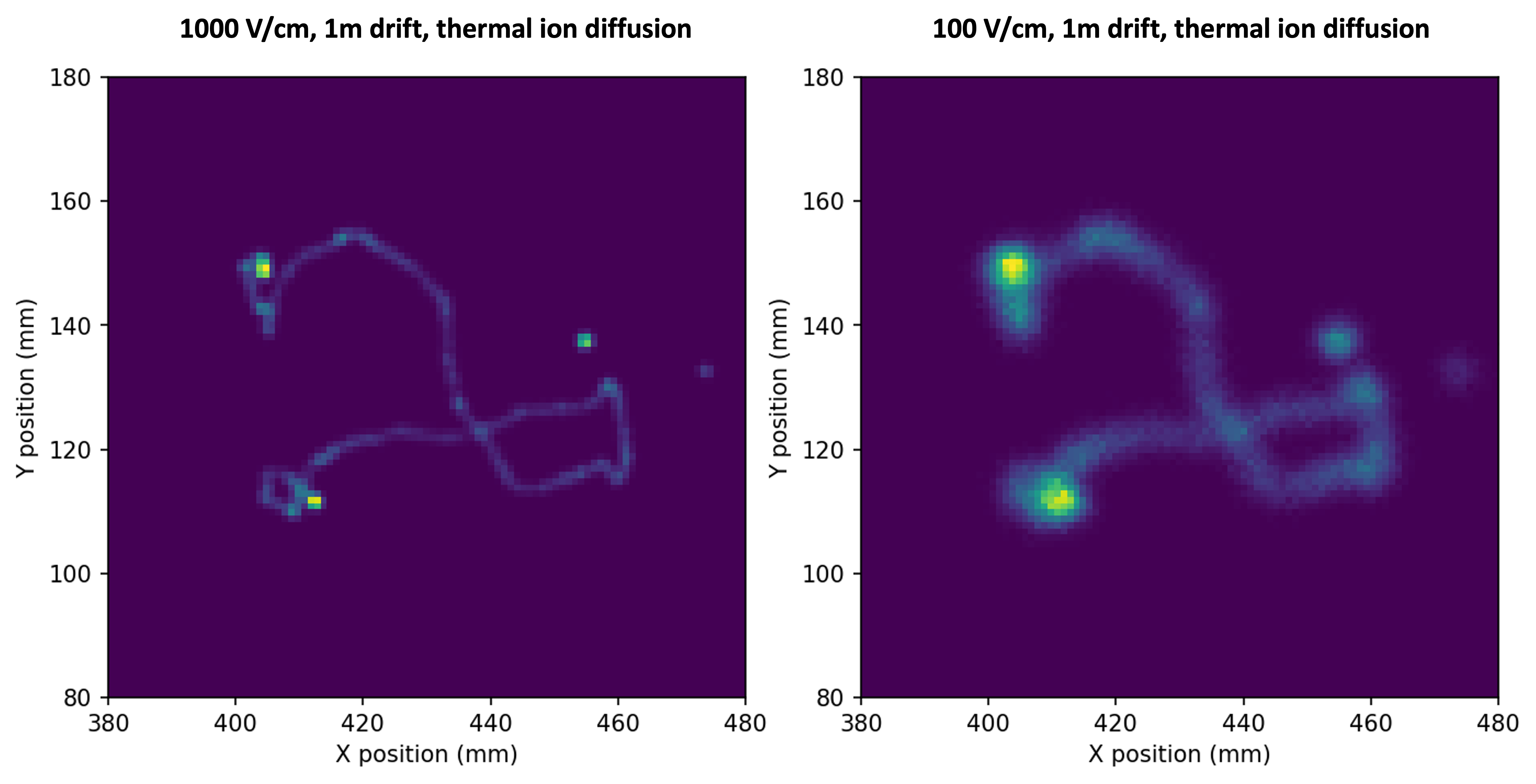} 

\caption{ A neutrinoless double beta decay event simulated at the Q-value for $^{82}$Se, drifted over 1~m and with thermal diffusion applied given an electric field of 1000~V/cm (left) and 100~V/cm (right). In both cases, very high resolution topological imaging appears possible by fluorescence imaging of the ion population arriving at the cathode. The two-electron topological signature is clearly seen in both cases. \label{fig:nubbimage}}
\end{center}
\end{figure}

To obtain a sense of the quality of imaging that may be achieved, Fig.~\ref{fig:nubbimage} shows the expected fluorescence intensity as a function of position on the anode plane for a simulated neutrinoless double beta decay event at the $^{82}$Se Q-value and 10~bar operating pressure. This event was simulated in GEANT4 in an SeF$_6$ gas volume at the appropriate density, and then thermal diffusion was applied as a random Gaussian displacement ion-by-ion assuming a 1~m drift distance and 1000~V/cm drift field (left) and 100~V/cm drift field (right), and 50\% of the arriving ions were assumed to lead to fluorescent turn-on behavior at the detection layer.  In both cases, exquisite topological imaging is possible, including clear identifiability of the end-points of the two electrons enabling topological identification.   It is notable that this improved transverse imaging comes at the cost of complete removal of longitudinal information, and so topological identification must thus be performed within a single projection. The extent to which this sacrifices efficiency or purity of the sample is a matter for a future study applying more advanced analysis methods.

\subsection{Other applications}

Other applications that mix the requirements of precision event reconstruction with large required masses may also be compelling use cases for a detector of this kind. Detection of $\nu_\tau$ via searches for sub-milimeter decay lengths of the $\tau$ lepton are one example. Such searches have typically been handled only with emulsion detectors in order to obtain the necessary vertexing resolution~\cite{kodama2008final,opera2014observation}, but a detector with fully active TPC-quality readout would naturally advance the study of $\nu_\tau$ and its oscillations.  Other applications involving precision track reconstruction that have motivated the use of gas-phase TPCs include studies of low-energy nucleon emission in neutrino interactions that have motivated proposal of a gaseous argon near detector for DUNE~\cite{singh2020quantifying,martin2017pressurized}.  Studies of rare eta decays~\cite{gatto2016redtop} and possible applications at a muon collider~\cite{mcdonald1999emittance} have also been suggested as strong applications for low density, high resolution TPCs, though notably these applications are imperfectly matched to this approach due to the high interaction rates involved.

\section{Conclusions}

In this work we have outlined a new detection concept: the Ion Fluorescent Chamber (IFC).  The IFC is envisioned as a solution for particle detection challenges that meet following three criteria:
\begin{enumerate}
    \item A physically very large detector is required;
    \item This detector must be imaged with precise spatial resolution over its entire volume;
    \item The event rate is low, so scanning after the event time is a plausible approach.
\end{enumerate}

Using one plane to trigger motion of a sensor to image fluorescence that is induced by ions of the opposite charge, ion-by-ion imaging at the thermal diffusion limit should be possible.  The trigger plane can be very coarsely instrumented but immediately responsive, whereas the imaging plane has resolution on the micron scale but a delayed readout.  This combination allows an in principle very large volume to be sensitized at a fine image resolution limited only by ion diffusion.  

We have suggested two seemingly natural applications that appear, to us, to stretch the capabilities of traditional large TPCs:  the directional reconstruction of nuclear recoils induced by dark matter that will be needed in order to pursue WIMP searches below the solar neutrino floor; and the search for neutrinoless double beta decay beyond the ton scale, where it may be necessary to realize extremely large detectors without compromising the performance parameters achievable in smaller systems.  Rudimentary explorations of both these applications suggest significant promise, though much quantitative work remains to be done to assess the true viability of this new approach.

\section*{Acknowledgements}
The University of Texas at Arlington physics group is supported by the Department of Energy under Early Career Award number DE-SC0019054 and Department of Energy Award DE-SC0019223, and the chemistry group by the National Science Foundation under award number NSF CHE 2004111, and the Robert A Welch Foundation, Y-2031-20200401.

\bibliographystyle{JHEP}
\bibliography{SeF6Whitepaper}

\end{document}